\documentstyle[epsfig]{europhys}

\def\And{{\rm and\ }}

\def\order{\mathop{\rm O}\nolimits}

\def\stars{\bigskip\centerline{***}\medskip}

\newif\ifboo \boofalse

\def\Review#1{\boofalse{\it #1},}
\def\Name#1{{\sc #1},}
\def\Vol#1{\ifboo Vol. {\bf #1}\else{\bf #1}\fi}
\def\Year#1{\ifboo #1\else(#1)\fi}

\def\Page#1{\ifboo {\rm p. #1}\else{\rm #1}\fi}

\def\gsim{\hbox{\lower3pt\vbox{\baselineskip=4pt \lineskiplimit=0pt \kern2pt 
          \hbox{$>$}\hbox{$\sim$}}}}
                                                                                
\euro{42}{3}{277}{1998}
\Date{1 March 1998}
\shorttitle{A. JASTER: ORIENTATIONAL ORDER OF THE TWO-DIMENSIONAL
HARD DISK SYSTEM}
                                                            
\title{Orientational order of the    
two-dimensional hard disk system}
 
\author{A. Jaster}
\institute{Universit\"{a}t - GH Siegen, D-57068 Siegen, Germany}      

\rec{12 February 1998}{17 March 1998}                            
\pacs{
\Pacs{64}{70.Dv}{Solid-liquid transitions}
\Pacs{64}{60.Fr}{Equilibrium properties near critical points, 
critical exponents}
\Pacs{61}{20.Ja}{Computer simulation of liquid structure}
}
                  
\begin{document}

\maketitle
\begin{abstract}
We report Monte Carlo results for the two-dimensional hard disk system.
Simulations were performed in the $NVT$ ensemble with up to 65536 
disks, using a new updating scheme. We analyze the 
bond orientational order 
parameter and correlation length in the isotropic phase and
the scaling behaviour of the bond orientational order parameter in the
transition region. The data are consistent with predictions
of the Kosterlitz-Thouless-Halperin-Nelson-Young theory, while a 
first-order phase transition is unlikely and a one-stage continuous 
transition 
can be ruled out.
\end{abstract}

The nature of the two-dimensional melting transition has been an unsolved
problem since many years \cite{STRAND,GLACLA}. The 
Kosterlitz-Thouless-Halperin-Nelson-Young (KTHNY) theory \cite{KTHNY}
predicts two continuous transitions. The first transition occurs
when the solid (quasi-long-range positional order, long-range
orientational order) undergoes a dislocation unbinding transition
to the hexatic phase (short-range positional order, quasi-long-range
orientational order). The second transition is the 
disclination unbinding transition which transforms this 
hexatic phase into an isotropic phase (short-range positional and 
orientational order). There are several other theoretical
approaches for the transition. One alternative   scenario
has been proposed by Chui \cite{CHUI}. He presented a theory via 
spontaneous generation of grain boundaries, {\it i.e.\/} collective excitations 
of dislocations, and predicted a conventional 
first-order phase transition from the solid to the isotropic
phase. In this case there exists a region where both phases coexist
instead of a hexatic phase.
Even for the simple hard disk system no consensus about the existence 
of a hexatic phase has been established. 

The melting transition of the hard disk system was 
first seen in a computer simulation by Alder and Wainwright 
\cite{ALDWAI}. They used a system of 870 disks and molecular
dynamics methods (constant volume $V$, energy $E$ and 
number of particles $N$ simulations)
and found that this system undergoes a first-order phase
transition. But the results of such small systems are
affected by large finite-size effects. Recent simulations used 
Monte Carlo (MC) techniques either with constant volume 
($NVT$ ensemble) \cite{ZOLCHE,WEMABI} or constant
pressure ($NpT$ ensemble) \cite{LEESTR,FEALST}. 
Lee and Strandburg \cite{LEESTR} 
used isobaric MC simulations and 
found a double-peaked structure in the volume distribution.
Lee-Kosterlitz scaling led them to conclude that the phase transition
is of first order. However, the data are not in the scaling region,
since their largest system contained only 400 particles.
MC investigations of the bond orientational order
parameter via finite-size scaling with the block analysis technique 
of 16384 particle systems were done by
Weber, Marx and Binder \cite{WEMABI}. They also  
favoured a first-order phase transition. In contrast to this, 
Fern\'{a}ndez, Alonso and Stankiewicz \cite{FEALST} 
predicted a one-stage continuous melting
transition, {\it i.e.\/}  a scenario without a hexatic phase.
Their conclusions were based on the examination of the bond orientational 
order parameter in very long runs of different systems up to 15876
particles and hard-crystalline wall boundary conditions.
The analysis of Zollweg and Chester \cite{ZOLCHE} for the pressure gave
an upper limit for a first-order phase transition, but is
compatible with all other scenarios.

In this letter, we present results obtained through MC simulations
in the $NVT$ ensemble to answer the question of the kind
of the phase transition.  
We consider systems of $N=32^2$, $64^2$,
$128^2$ and $256^2$ hard disks in a two-dimensional square 
box. We find that finite-size effects with these
boundary conditions are 
not substantially larger than in a rectangular box with ratio 
$\sqrt{3}:2$, furthermore no simulations in the solid phase were made.
The disk diameter is set equal to one.  For the simulations a
new updating scheme was developed 
\cite{JASTER1}, in which the conventional Metropolis step of a single
particle is replaced by a collective (non-local) step of a chain of particles.
A cell structure was 
chosen such  that one cell can only be occupied by a single
disk. In all updatings, the random number generator proposed 
by L\"{u}scher \cite{LUSCHER} was applied. The simulations
were performed on a Silicon Graphics workstation and a CRAY
T3E. In the latter case, we used the different nodes of the
parallel machine to generate independent data sets.
Statistical errors have been calculated by binning. 
Additionally, we performed a jackknife analysis of the different
data sets from the different nodes. 
Careful attention has been paid to the equilibration of all systems.
For example, we performed $3\times 10^5$ 
`sweeps' for $N=256^2$ at $\rho=0.890$
with the improved (chain) Metropolis updating scheme to warm up the system
and $1.9 \times 10^6$ `sweeps' to measure the expectation values
(for 6 independent data sets).
The acceptance rate for this run was about $54\%$.
Further details will  be published later \cite{JASTER2}.

Simulations were performed in the isotropic phase and in the
phase transition region. 
In the isotropic phase we measured the (global) bond orientational order
parameter $\psi_6$ and the correlation length of the bond
orientation $\xi_6$.
The local value of $\psi_6$
for a particle $i$ located at ${\rm \bf x}=(x,y)$ is given by
\begin{equation}
\psi_6({\rm \bf x})=
\frac{1}{N_i} \sum_j \exp \left (6\, {\rm i} \, \theta_{ij} \right ) \ ,
\end{equation}
where the sum on $j$ is over the $N_i$ neighbours of this particle
and $\theta_{ij}$ is the angle  between the particles $i$
and $j$  and an arbitrary but fixed reference 
axis. Neighbours are obtained in a
usual way by the Voronoi construction. The (global) bond orientational order
parameter is just the absolute value of the  average over all particles:
\begin{equation}
\psi_6= \left | \frac{1}{N} \sum \psi_6({\rm \bf x})  \right | \ . 
\end{equation}
The bond orientational correlation length
was extracted from the `zero-momentum' correlation function
of $\psi_6({\rm \bf x})$
\begin{equation}
g_6(x) = \left \langle  \left (
\frac{1}{N_k} \sum_y \psi_6(x,y)
\right ) ^* \, \left (
\frac{1}{N_k'} \sum_{y'} \psi_6(0,y')
\right ) \right \rangle \ ,
\end{equation}
by fitting the data with a single $\cosh$, where 
$N_k$ denotes the number of particles in a stripe between
$x+\Delta x/2$ and $x-\Delta x/2$.
This method allows a precise  measurement
of $\psi_6$ apart from some systematical errors, which will
be large --- compared to our statistical errors ---
for small values of $\xi_6$
(for details see \cite{JASTER2}).
Additionally, we calculated the radial bond 
orientational correlation function
\begin{equation}
g_6(r)= \langle {\psi_6}^*(0) \, \psi_6(r)
\rangle / g(r) \ ,
\end{equation}
and extracted the correlation length from an {\it Ansatz} of the form
$g_6(r) \sim r^{-1/2} \exp(-r/\xi_6)$, where 
$g(r)$ is the pair correlation function.
In all simulations of the isotropic phase we
used systems of at least $N=64^2$ particles
and chose the box length 
to satisfy $L>7\xi_6$. 
In these cases, within statistical errors, we found  no finite-size effects 
on the bond orientational correlation length and on the 
susceptibility 
\begin{equation}
\label{eq_sus}
\chi_6=N \langle {\psi_6}^2 \rangle \ .
\end{equation}
Equation (\ref{eq_sus}) differs from  
$\chi_6=N (\langle {\psi_6}^2 \rangle -\langle \psi_6 \rangle ^2 )$
by a factor $1-2/\pi$ in the thermodynamic limit.
Due to the new updating scheme, which reduces the autocorrelation time, 
we were able to perform simulations with large correlation lengths,
namely close to the
disclination binding transition point $\rho_{\rm i}$. 

The KTHNY scenario predicts an exponential singularity for
the correlation length 
\begin{equation}
\label{KTxi}
\xi_6(t) = a_{\xi} \, \exp \left ( b_{\xi}\, t^{-1/2}  \right ) \ , 
\end{equation}
and the susceptibility  $\chi_6$
\begin{equation}
\label{KTchi}
\chi_6(t) = a_{\chi} \, \exp \left ( b_{\chi}\, t^{-1/2} \right )
\end{equation}
if $t=\rho_{\rm i} -\rho \rightarrow 0^+$.
The critical exponent $\eta_6$
defined by
\begin{equation}
\label{chixi}
\chi_6 \sim {\xi_6}^{2-\eta_6} \ ,
\end{equation}
is given by $ \eta_6=1/4$,
while  $b_\xi$  is a non-universal constant and 
\begin{equation}
\label{etabb}
b_\chi=(2-\eta_6)b_\xi \ .
\end{equation}
\begin{figure}[t]
\centerline{\epsfxsize=8.5cm \epsfbox{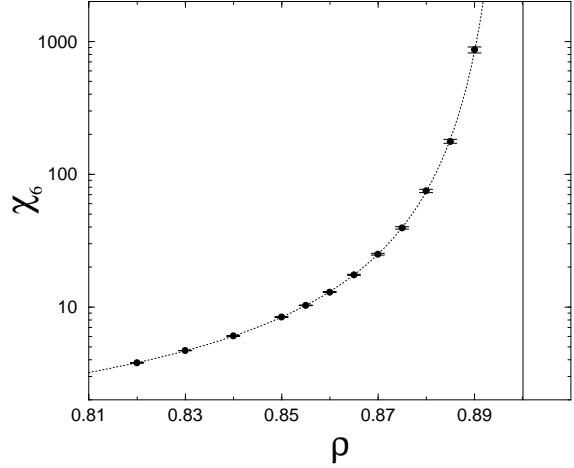}}
\caption{\label{fig_sus}
Susceptibility as a function of the density. The curve shown
is the best fit for a KTHNY behaviour. The critical value of
$\rho$ is visualized by a vertical line.}
\end{figure}

\begin{figure}[t]
\centerline{\epsfxsize=8.5cm \epsfbox{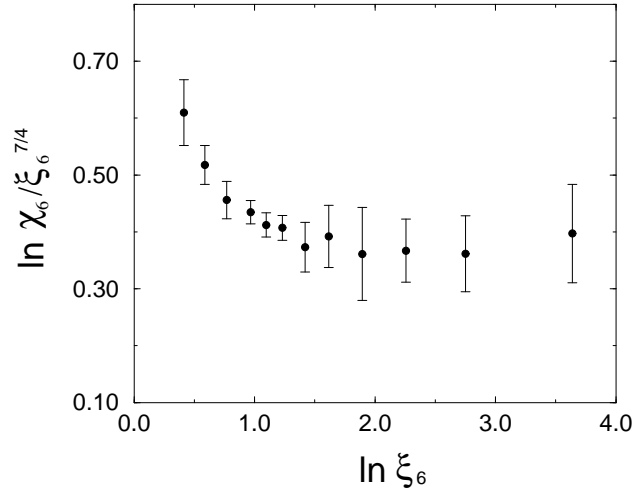}}
\caption{\label{fig_corsus}
Test of the scaling relation $\chi_6 \sim \xi_6^{7/4}$. The slope 
gives the deviation from $\eta_6=1/4$.}
\end{figure}

We analyzed the critical behaviour of $\xi_6$ and $\chi_6$
by performing least square fits according to eqs.\
(\ref{KTxi}) and (\ref{KTchi}).
Typically, statistical errors of $\xi_6$ and $\chi_6$
are in the range $1\%-5\%$.
Errors for the fitting parameter were computed 
by performing fits on data sets being Gaussian
distributed around the expectation value. 
If we used all 12 different measurement points with
$0.82 \leq \rho \leq 0.89$  we got a
$\chi^2$ per degree of freedom (d.o.f.) of $0.75$ 
for $\xi_6(t)$ and $0.65$ for $\chi_6(t)$, {\it i.e.\/}
the data are in a very good agreement with
an exponential singularity of the KTHNY type.
The critical values of $\rho$ were given by
$\rho_{\rm i}=0.9017(7)$ and $0.9002(3)$,
respectively. The results for the susceptibility are
shown in  fig.~\ref{fig_sus}. 
Data far away from the transition (in particular  
for the correlation length) are affected by 
systematical errors.  
Therefore, fits were also performed omitting some data
at low densities. For example, for the eight points      
with $0.855 \leq \rho \leq 0.89$ (for $\xi_6 \gsim \, 3.0$) 
we got $\chi^2/$d.o.f.$=0.23$,
$\rho_{\rm i}=0.9006(8)$ for $\xi_6(t)$ and
$\chi^2/$d.o.f.$=0.58$, $\rho_{\rm i}=0.9000(4)$ for $\chi_6(t)$.
The critical exponent $\eta_6$ is calculated using 
eq.~(\ref{etabb}), yielding $\eta_6=0.451(21)$ and $0.349(44)$,
respectively. A detailed analysis shows that the value of $\eta_6$
decreases if $t \rightarrow 0^+$. This can be seen, if we use
eq.~(\ref{chixi}) and plot $\ln(\chi_6/{\xi_6}^{7/4})$ versus $\ln(\xi_6)$.
For the predicted value $\eta_6=1/4$ we should see a horizontal line.
A different value of $\eta_6$ would correspond to a straight line with
a non-zero slope.  Indeed, there is  a negative
slope for small values of $\xi_6$ as can be seen in fig.~\ref{fig_corsus}.  
Nevertheless, in the limit $\xi_6 \rightarrow \infty$ 
the data are compatible with $\eta_6=1/4$. A fit with the last six
data points gives $\eta_6=0.251(36)$.

\begin{figure}[t]
\centerline{\epsfxsize=8.5cm \epsfbox{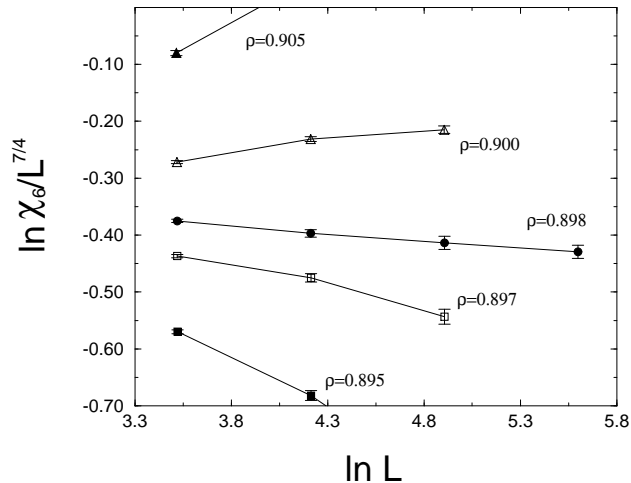}}
\caption{\label{fig_FSSchi}
Finite-size scaling of the susceptibility for various densities.
The slope gives the deviation from $\eta_6=1/4$. The lines are guides
to the eye.}
\end{figure}

We now come to the simulations with $\rho \approx \rho_{\rm i}$.
Finite-size scaling (FSS) implies $\chi_6 \sim L^{2-\eta_6}$
at $\rho=\rho_{\rm i}$ for large enough systems.
For $\rho < \rho_{\rm i}$, 
corrections for finite correlation lengths
of $\order (L/\xi_6)$ have to be taken  into account. For 
$\rho_{\rm i} < \rho \le \rho_{\rm m}$,
$\eta_6$ is a function of the density. As $\rho$ approaches 
the melting density
$\rho_{\rm m}$, {\it i.e.\/} at the end
of the hexatic phase, $\eta_6 \rightarrow 0^+$.
Figure \ref{fig_FSSchi} shows $\ln(\chi_6/L^{7/4})$ versus 
$\ln(L)$ for various $\rho$.
The slope was extracted from linear
fits and gives the deviation from $\eta_6=1/4$. 
Using the FSS behaviour to locate $\rho_{\rm i}$, 
the requirement $\eta_6(\rho_{\rm i})=1/4$ yields 
$\rho_{\rm i}=0.899(1)$. This value is in agreement with that 
obtained from the singularities of $\xi_6(t)$ and $\chi_6(t)$.
A slightly different value of $\eta_6$ would not alter this situation.
Moreover, our estimates of $\rho_{\rm i}$ agree with Weber {\it et al.\/}
\cite{WEMABI} who used the fourth-order cumulant intersection
($\rho_{\rm i}=0.8985(5)$). However, it differs from  their
value  obtained using the singularity of $\chi_6$ ($\rho_{\rm i}=0.913$).
The result $\rho_{\rm i}=0.916(4)$ of Fern\'{a}ndez {\it et al.\/} \cite{FEALST} 
is not compatible with our value.

Our MC data suggest that the the orientational order behaves
as predicted by the KTHNY theory. A one-stage continuous transition
\cite{FEALST} can be ruled out, since  
$\rho_{\rm m}  \gsim \, 0.910$  (obtained 
from $\eta(\rho_{\rm m})=0$) is away from 
$\rho_{\rm i}$ in this work. Also our data are not compatible with a
first-order phase transition with small correlation lengths,
since we find no deviation from the predicted 
singularities of $\chi_6$ and
$\xi_6$ up to $\xi_6 \approx 38$.
(Alternative approaches for the singularities result
in large $\chi^2/$d.o.f. For example, a conventional second-order
behaviour with a power-law singularity
of the form $\ln(\xi_6)=a-\nu \ln(t)$ yields $\chi^2/$d.o.f.=$4.1$.)
We also have examined FSS of the
fourth-order cumulant 
$U=1- \langle {\psi_6}^4\rangle/3 \langle {\psi_6}^2\rangle^2$.
In the hexatic phase, FSS implies scale invariance of $U$.
The narrowness of such a scale invariant region
was one argument in ref.\ \cite{WEMABI} against the existence
of a hexatic phase. Unfortunately, statistical errors in
our data are too large to answer this question.
Although our data cannot rule out a first-order phase transition with 
very large orientational order correlation lengths, a KTHNY-like phase 
transition seems to be more likely.

\stars

We thank H.~Hahn for helpful discussions and the 
Institute of Scientific Computing in  Braunschweig
for providing computer time on their CRAY T3E.
 


\end{document}